# PhyloFrame: A DataFrame-based Library for Fast, Flexible Phylogenetic Computation

**Matthew Andres Moreno** [1,2,3,4], **Jeet Sukumaran** [5,6], **Luis Zaman** [1,2,4], **and Emily Dolson** [7,8,9]

**1** Department of Ecology and Evolutionary Biology **2** Center for the Study of Complex Systems **3** Michigan Institute for Data and AI in Society **4** University of Michigan, Ann Arbor, MI, United States of America **5** Department of Biology **6** San Diego State University, San Diego, CA, United States of America **7** Department of Computer Science and Engineering **8** Program in Ecology, Evolution, and Behavior **9** Michigan State University, East Lansing, MI, United States of America

## Summary

PhyloFrame is a Python library for phylogenetic computation targeting the gap between specialist, compiler-optimized operations and fexible, script-based workfows — with emphasis on fast, memory-efcient operations for very large tree sizes (e.g., $\geq$ 300,000 taxa).

PhyloFrame is built around a DataFrame-based tree representation, where each row corresponds to a node and columns record ancestor relationships, branch lengths, taxon labels, and any user-defned attributes. Crucial for scalability, such array-backed storage allows both library and end-user code alike to seamlessly harness Just-in-Time (JIT) compilation (e.g., Numba) and vectorized execution (e.g., NumPy, Polars). At large tree sizes, performance generally matches or exceeds Python libraries backed by native code — notably, achieving strong performance in topological-order traversals and Newick I/O.

DataFrame-based representation afords several additional conveniences, including:

- succinct bulk operations (e.g., NumPy);
- powerful queries and transformations (e.g., Polars expressions, Pandas indexing, SQL-style joins and merges);
- compatibility with modern tabular data formats that are compression-friendly, type-aware, nullable, and highly portable (e.g., Parquet); and
- broad interoperation with table-oriented data science tools (e.g., Seaborn, Plotly, Vega-Altair, tidyverse, Excel).

Current library features include tree input/output, synthetic tree generation, taxon-based queries, tree traversals, tree metrics, tree manipulation, tree downsampling, and tree comparison. Most functionality supports both Pandas and Polars DataFrames, and is available through programmatic and CLI-based interfaces.

## Statement of Need

In addition to the ever-growing infux of high-throughput sequence data ( Stephens et al., 2015), recent years have seen the advent of powerful biotechnologies for cell-lineage tracing (McKenna et al., 2016; Nguyen Ba et al., 2019), ultra-high-throughput workfows capable of estimating ancestry among hundreds of millions of taxa (Konno et al., 2022), and ultra-scale simulations generating billion-taxa lineage histories (Singhvi et al., 2025). These emerging sources of phylogeny data (evolutionary trees) ofer unprecedented visibility into developmental and eco-evolutionary processes (Chan et al., 2019; Faith, 1992; French et al., 2023; Kim et al., 2006; Lenski et al., 2003; Lewinsohn et al., 2023; Nozoe et al., 2017; Stamatakis, 2005) — but data volume far exceeds the capabilities of traditional tools supporting phylogenetic workfow development (Cock et al., 2009; Huerta-Cepas et al., 2016; Moreno, Holder, et al., 2024).

This unmet need has prompted development of several performance-frst libraries for phylogenetic computing in Python (Moshiri, 2020, 2025; Neches & Scott, 2018). These libraries have greatly improved the state of the art by providing memory-efcient data structures and powerful compiler-optimized implementations of standard phylogeny-based calculations. End-user code requiring custom iterative operations, however, typically remains as slower, interpreted Python bytecode. Achieving full compiler optimization for custom operations requires integration of less familiar systems-level code (e.g., C/C++).

PhyloFrame seeks to complement existing software by flling the gap between fully-optimized native code and custom end-user operations. For this purpose, PhyloFrame sacrifces a typical object-oriented interface in favor of a DataFrame-based representation of tree nodes. In this framework, a tree with $n$ nodes boils down to a set of named length-$n$ arrays — each storing a particular node-attribute (e.g., ancestor index, origin time, taxon name, etc.). This array-backing enables vectorized bulk operations (e.g., NumPy (Harris et al., 2020)) and JIT compilation (e.g., Numba (Lam et al., 2015)) — which allow competitive performance, while retaining Python-level expressiveness and productivity.

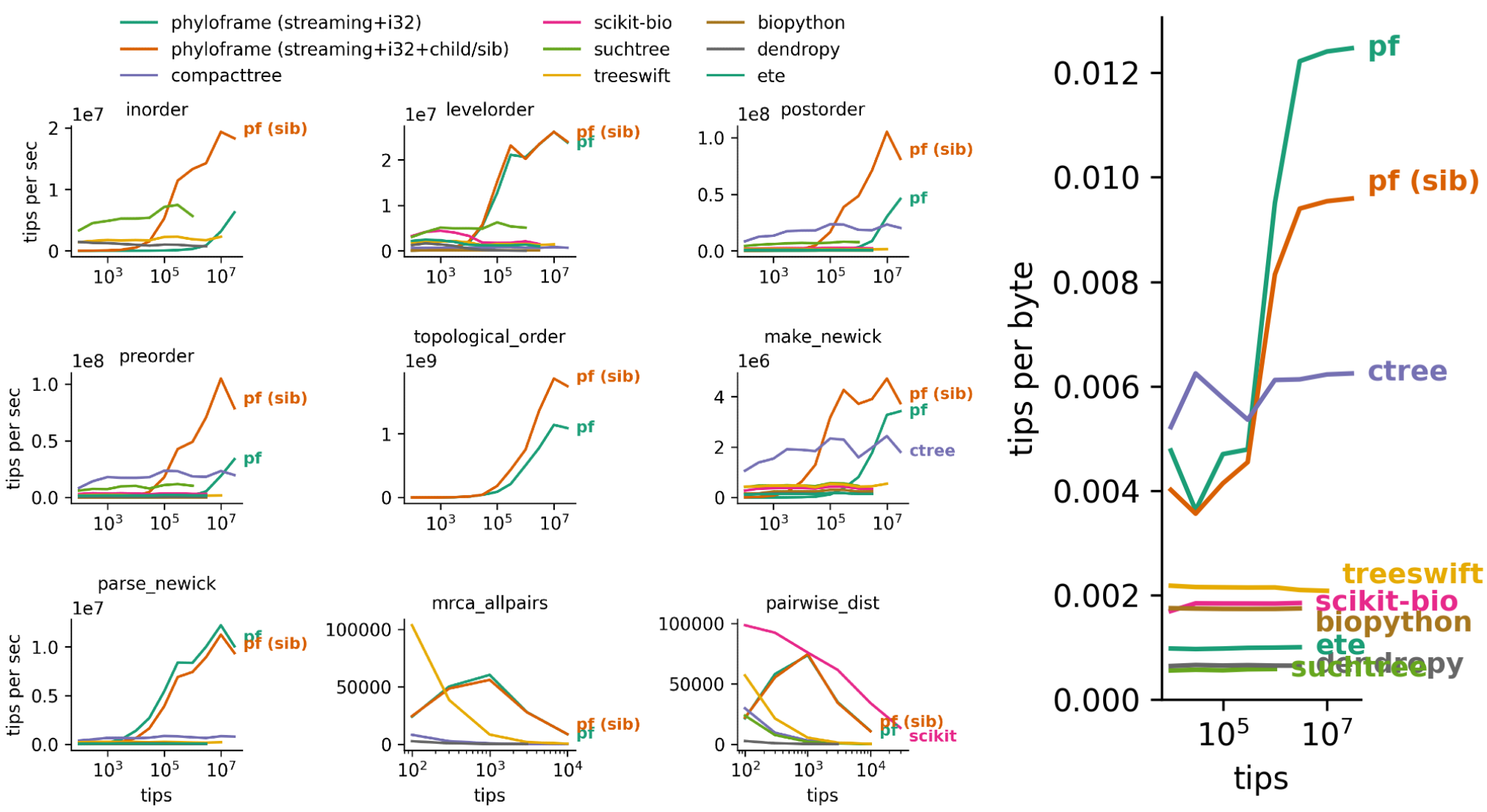

**Figure 1:** Benchmark comparison. Left: throughput (tips processed per second) by operation across tree sizes. Right: memory efciency (tips stored per byte), across tree sizes. Higher is better in both panels.

Figure 1 benchmarks throughput and memory efciency for Python-based operations on balanced binary trees with up to 30 million tips.[1]

Beyond tree sizes of around 300,000 tips[2], PhyloFrame matches the throughput and efciency of native-backed libraries (e.g., CompactTree, SuchTree) for most benchmarked operations. At very large tree sizes (e.g., $\geq 1$ million tips), PhyloFrame substantially accelerates throughput for some operations. For traversals, benefit likely stems from capability to materialize node iteration within a JIT-compiled context. Newick parsing, on the other hand, likely benefits from streamlined per-array memory allocation (as opposed to allocating node objects or setting up child lists), while Newick generation leverages the Polars engine to accelerate string-building.

[1]Timings were conducted on GitHub action `ubuntu-24.04` runners (4-core x86/16GB memory circa May 2026), with cross-library comparisons restricted to common job instances for parity. Packages were installed via pip, `phyloframe[jit]==0.10.0`, `biopython==1.86`, `CompactTree==1.0.0`, `dendropy==5.0.8`, `treeswift==1.1.45`, `ete3==3.1.3`, `scikit-bio==0.6.3`, and `SuchTree==1.3`. PhyloFrame benchmarks report Polars-based operations. Raw data is archived at https://osf.io/knw8x (Foster & Deardorf, 2017 ). Benchmark design follows (Moshiri, 2025).

[2]For workloads involving large quantities of smaller trees, performance beneft can be achieved by consolidating data as a "forest" within a single DataFrame.

PhyloFrame's architecture most closely resembles CompactTree, but difers in that taxon label and branch length attributes are optional, arbitrary end-user attributes are supported directly, and child lists use a fat array rather than nested structures (e.g., via compressed sparse row or linked lists). Under both libraries, topological-order traversals are particularly efcient, as they correspond to a sequential scan over array memory. Importantly, CompactTree's design priority is in supporting C++-based usage, which is strongly recommended over its Python-based bindings for performance — a scenario not captured by the Python-based benchmarks reported here.

An important performance trade-of not captured in these benchmarks is tree manipulation. While DataFrame-based representation does support mutable tree reconfguration after construction, unlike allocated node-and-pointer representations, one-of node creation and deletion is not guaranteed $\mathcal{O}(1)$.

Beyond high-performance end-user extensibility, DataFrame-based representation afords several useful conveniences that complement the existing phylogenetic computing landscape — briefy reviewed in Appendix A.

## Features

PhyloFrame supports the following operations on both Pandas and Polars DataFrames (McKinney, 2010; Vink et al., 2024):

- **tree input/output**: Newick and ALife Data Standard formats (Lalejini et al., 2019);
- **tree synthesis**: structured (e.g., comb, balanced, star) and random (e.g., edge-adding, node-adding);
- **taxon-based queries**: pairwise and all-pairs MRCA/patristic distance;
- **tree traversals**: preorder, postorder, inorder, levelorder, semiorder, topological;
- **tree metrics**: Colless imbalance (Colless & Wiley, 1982), Sackin index (Sackin, 1972), Faith's phylogenetic diversity (Faith, 1992);
- **tree manipulation**: collapsing unifurcations, rerooting, ladderizing;
- **tree downsampling**: lineage sampling, tip sampling, clade sampling, custom pruning;
- **tree comparison**: triplet/quartet distance (Dobson, 1975; Estabrook et al., 1985; Sand et al., 2014) and topological isomorphism; and
- **tree visualization**: integration with Matplotlib-based iplotx (Zanini, 2025) and interactive Taxonium web app[3] (Sanderson, 2022).

A quickstart guide and full API listing are included in the PhyloFrame documentation at https://phyloframe.readthedocs.io. PhyloFrame is installable from the Python Package Index (PyPI) via `pip` (e.g., `python3 -m pip install "phyloframe[jit]"`).

## Demo: End-user JIT Compilation and Tidy Plotting

Example code uses a pipeline pattern to (1) generate a 100-tip tree, (2) apply JIT-compiled end-user code to simulate trait inheritance, (3) mark attributes `node_depth` and `is_leaf`, and (4) visualize using Seaborn and Matplotlib.

```
import numpy as np; import polars as pl; import seaborn as sns
from phyloframe import _auxlib as pfa
from phyloframe import legacy as pfl
```

[3] via an experimental fork at https://mmore500.github.io/taxonium.

```python
@pfa.jit(cache=False, nopython=True)  # JIT compile via Numba, if available
def simulate_trait(ancestor_ids: np.ndarray) -> np.ndarray:
    trait = np.zeros(ancestor_ids.size, dtype=float)
    for id_, anc_id in enumerate(ancestor_ids):
        if id_ == anc_id: continue  # exclude root
        trait[id_] = trait[anc_id] + np.random.normal()
    return trait

def plot_trait(data: pl.DataFrame) -> None:
    selectors = dict(x="node_depth", y="trait", hue="is_leaf")
    style = dict(legend=False, palette=["gray", "steelblue"], s=15)
    ax = sns.scatterplot(data, **selectors, **style)

    depth, ancestor, trait = data[["node_depth", "ancestor_id", "trait"]]
    segments = [[depth, depth[ancestor]], [trait, trait[ancestor]]]
    ax.plot(*segments, color="#CCCCCCCC", zorder=-2)  # link parent/child

pfl.alifestd_make_edge_split_polars(n_leaves=100, seed=42,  # random tree
    ).pipe(pfl.alifestd_topological_sort_polars,  # parents before children
    ).pipe(pfl.alifestd_assign_contiguous_ids_polars,  # reassign ids
    ).pipe(pfl.alifestd_mark_node_depth_polars,  # add node_depth col
    ).pipe(pfl.alifestd_mark_leaves_polars,  # add is_leaf col
    ).with_columns(trait=pl.col("ancestor_id").map_batches(  # add trait col
        lambda x: simulate_trait(x.to_numpy()), return_dtype=float,
)).pipe(plot_trait)  # draw tree
```

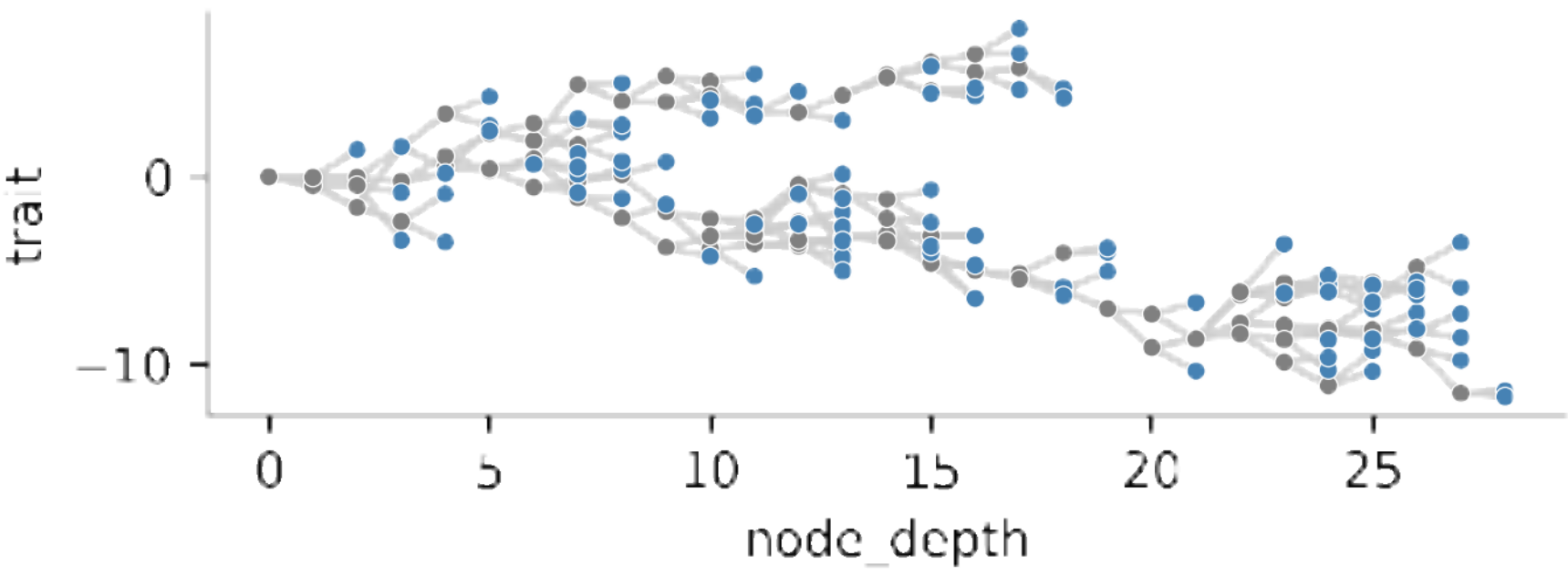

**Figure 2:** Trait simulation visualization

## Related Software

A rich ensemble of existing libraries supports Python-based phylogenetic computing.

- DendroPy (Moreno, Holder, et al., 2024) ofers a comprehensive and dependency-free object-oriented framework for phylogenetic simulation and analysis.
- Biopython (Cock et al., 2009; Talevich et al., 2012) includes a Phylo module with a tree data structure and general-purpose utilites (including visualization).
- ETE (Huerta-Cepas et al., 2016) combines tree reconstruction, analysis, and visualization capabilities, alongside integration with the NCBI taxonomy database.
- scikit-bio (Aton et al., 2026) provides a broad bioinformatics toolkit, including tree manipulation, reconstruction, and phylogenetic diversity metrics.
- tskit (Kelleher et al., 2016; Wong et al., 2024) implements the succinct tree sequence data structure, a compact representation of correlated local trees along a recombining genome, used to represent Ancestral Recombination Graphs.

- CompactTree (Moshiri, 2025) provides a memory-efcient array-based data structure implementation, distributed as a C++ header-only library with a Python wrapper.
- TreeSwift (Moshiri, 2020) is a performance-oriented pure-Python library using a compact linked-node representation, designed to support very large trees.
- SuchTree (Neches & Scott, 2018) uses a Cython-based array data structure, focusing on fast pairwise distance queries and co-phylogenetic analyses; operations release the Python GIL (Global Interpreter Lock) to allow multithread parallelism.
- ToyTree (Eaton, 2019) is an object-oriented library, providing integrated visualization functionality.
- PhyloTrack (Dolson et al., 2024) uses a node-and-pointer data structure to record lineage histories from forward-time agent-based models, with support for on-the-fy extinction pruning and metric calculations.

In the Julia ecosystem, PhyloNetworks (Solís-Lemus et al., 2017) emphasizes fexible support for generalized phylogenetic networks incorporating reticulation events. The R-based ecosystem has largely coalesced around ape's edge matrix tree representation (Paradis & Schliep, 2018). Other work has applied graph databases to manage large-scale phylogeny data (Lourenço et al., 2024).

Except for commonality in name, the PhyloFrame library presented here is unrelated to recent machine learning methodology developed to counteract ancestral bias in precision medicine (Smith et al., 2025).

## Projects Using the Software

PhyloFrame originated from phylogeny utilities developed within the hstrat library for phylogeny tracking in highly-distributed evolution simulations (Moreno et al., 2022b). PhyloFrame utilities have contributed substantially to several projects, including analysis of large-scale simulations scaling up to a billion taxa (Moreno, Ranjan, et al., 2025; Moreno, Yang, et al., 2024; Moreno et al., 2022a; Moreno, Rodriguez-Papa, et al., 2025; Singhvi et al., 2025).

## Development Roadmap

Much future work remains in development of the PhyloFrame library.

Feature-level improvements (e.g., tree metrics, manipulations) are planned on an as-needed basis, with requests welcome via the project issue tracker at https://github.com/mmore500/phyloframe/issues.

Looking further ahead, a redesigned API is planned to accompany PhyloFrame's v1 release. In anticipation of this release, all current PhyloFrame operations are packaged in `phyloframe.legacy`. This API is stable and will continue to be maintained for backward compatibility.

Identifed design and development priorities include:

- better-standardized naming schemes for library-function-generated columns,
- greater API symmetry between Pandas and Polars functionality,
- automatic cleanup of columns invalidated by tree manipulation,
- automatic repair of tree invariants (i.e., contiguous ids, topological order),
- automatic caching of tree invariant safety checks (currently, bypassable via environment variable),
- frst-class support for reticulated ancestry graphs and unrooted trees,
- support for GPU-based computations via CuPy and the RAPIDS Pandas integration (Okuta et al., 2017; Team, 2023),

- wheel-based distribution of pre-compiled Numba artifacts, and
- high-level visualization utilities leveraging Seaborn and iplotx (Waskom, 2021; Zanini, 2025).

Beyond Python, DataFrame-based phylogenetic computing may prove useful in other programming language ecosystems, such as Julia and R. Julia appears especially well-suited, on account of its tightly-integrated DataFrame stack (Bouchet-Valat & Kamiński, 2023) and built-in support for JIT compilation (Bezanson et al., 2017).

# Acknowledgements

This material is based upon work supported by the Eric and Wendy Schmidt AI in Science Postdoctoral Fellowship, a Schmidt Sciences program. This material is based upon work supported by the U.S. Department of Energy, Ofce of Science, Ofce of Advanced Scientifc Computing Research (ASCR), under Award Number DE-SC0025634. This report was prepared as an account of work sponsored by an agency of the United States Government. Neither the United States Government nor any agency thereof, nor any of their employees, makes any warranty, express or implied, or assumes any legal liability or responsibility for the accuracy, completeness, or usefulness of any information, apparatus, product, or process disclosed, or represents that its use would not infringe privately owned rights. Reference herein to any specifc commercial product, process, or service by trade name, trademark, manufacturer, or otherwise does not necessarily constitute or imply its endorsement, recommendation, or favoring by the United States Government or any agency thereof. The views and opinions of authors expressed herein do not necessarily state or refect those of the United States Government or any agency thereof.

**AI Use Declaration:** During the preparation of this work, AI tools were used to assemble manuscript boilerplate, proofread manuscript drafts, and assemble benchmarking scripts. Library maintenance employs agentic software development to refactor code, draft documentation, and implement scoped library features. Such contributions are closely supervised and tracked via commit co-authorship. Tools used include Claude Code, Google Gemini, and OpenAI ChatGPT.

## Appendix A: Why a DataFrame-based Tree Representation?

PhyloFrame relies on a fully-tabular data structure hosted within DataFrame objects (e.g., `pandas.DataFrame`, `polars.LazyFrame`, `polars.DataFrame`, etc.). Because DataFrame-based computation provides the foundation for much of modern data science (Wu, 2020), a rich infrastructure ecosystem has developed around it — providing several notable synergies to phylogenetic computation workfows.

**Fast and highly portable load/save.** DataFrame libraries provide robust builtin I/O capabilities (e.g., `pandas.read_csv`, `polars.read_parquet`, R's `read.table`). Many implementations can automatically fetch from URLs, cloud providers (e.g., AWS S3, Google Cloud, etc.), and online repositories (Foster & Deardorf, 2017 ; Singh, 2011). Contiguous allocations allow fast tree deserialization (e.g., Newick) and tree generation.

**Beneft from modern tabular fle formats.** Granular deserialization of selected columns, columnar compression for efcient storage, categorical strings, and explicit column typing with frst-class null representation (e.g., Parquet ( Vohra, 2016)). Data layout optimization for fast load/save (e.g., Feather (Wickham & McKinney, 2016)).

**Beneft from modern high-performance DataFrame tooling.** Memory-efcient representation, larger-than-memory streaming operations (e.g., Polars), distributed computing operations (e.g., Dask (Rocklin, 2015)), multithreaded operations (e.g., Polars), vectorized operations (e.g., NumPy), and just-in-time compilation (e.g., Numba).

**Beneft from rich, expressive DataFrame functionality.** Leverage powerful querying and transformation APIs (e.g., Polars expressions, Pandas indexing) for fexible fltering, bulk calculations, grouped aggregations, join/merge operations, and chained transformations directly over tree data without manual loops.

**Cache-friendly, memory-efcient, fexible data structure.** Data occupies contiguous arrays, expediting tree creation and topological order traversals (e.g., parents before children or vice versa). Base memory footprint is lightweight (e.g., as little as 32 bits per node), but can be dynamically augmented to expedite traversals and calculations (e.g., "linked list" of children via columns for frst child/next sibling indices).

**Interoperation.** Multi-language interoperation (e.g., possible future support for zero-copy interop between R and Python via reticulate and Arrow (Developers, 2024; Ushey et al., 2026), possible future support for zero-copy Polars DataFrames shared between Rust and Python). Multi-library interoperation (e.g., highly-optimized or zero-copy interoperation between Polars and Pandas; Python DataFrame protocol (Meurer, 2023)). Interoperation with broader Python DataFrame ecosystem (developers, 2024; Vallat, 2018; VanderPlas et al., 2018; Waskom, 2021) and ALife data standard tooling (Lalejini et al., 2019).

## Appendix B: Tree Manipulation Pipeline Demo

Example code shows sequential tree transforms applied using a pipeline pattern. Such complex tree manipulations, including custom operations, can often be performed succinctly without loops or recursion.

```
import numpy as np; from pandas import DataFrame
from phyloframe import legacy as pfl

df_raw: DataFrame = pfl.alifestd_from_newick("(((r:1)c:2,(x:2,y:1)e:1.5)s:2)a;")
df_res: DataFrame = df_raw.drop(columns=["branch_length", "origin_time_delta"],
    ).pipe(pfl.alifestd_reroot_at_id_asexual,  # reroot at node "r"
      new_root_id=df_raw.query("taxon_label == 'r'")["id"].item(),
    ).pipe(lambda df: df.assign(branch_length=np.where(  # flip rerooted lengths
```

```
        df_raw.loc[df["id"], "ancestor_id"] != df["ancestor_id"],  # where flipped
        df_raw.loc[df["ancestor_id"], "branch_length"],  # take ancestor's value
        df_raw.loc[df["id"], "branch_length"],  # ...otherwise keep own
    ))).pipe(pfl.alifestd_to_working_format,  # reassign id values
    ).pipe(pfl.alifestd_mark_lineage_cumsum_asexual,  # accumulate branch length
        mark_as="origin_time", values="branch_length",  # ...to mark origin time
    ).pipe(pfl.alifestd_sort_children_asexual, criterion="taxon_label",
    ).pipe(pfl.alifestd_to_working_format,  # reassign id values
    ).pipe(pfl.alifestd_ultrametricize, method="extend")  # align tip times
```

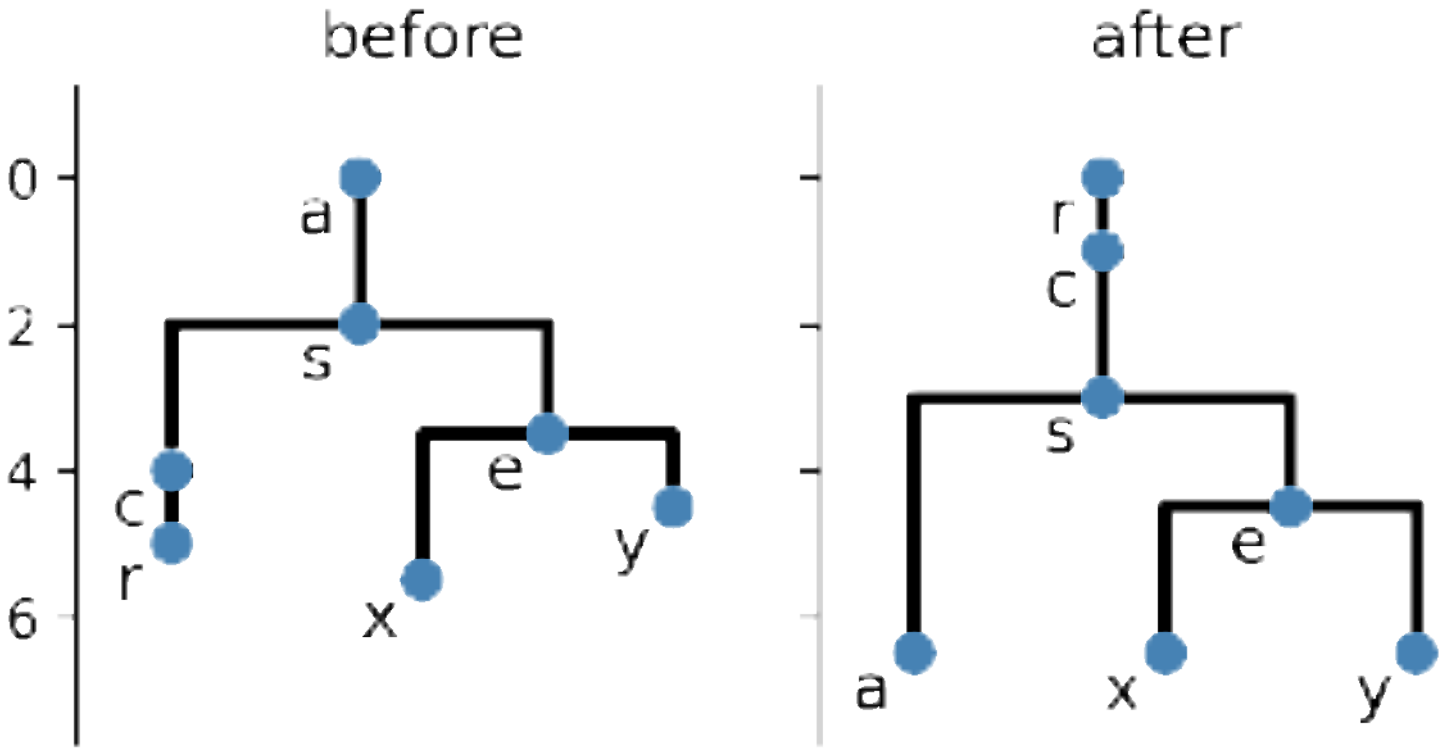

**Figure 3:** Before and after tree plots, rendered via integration with iplotx.

## Appendix C: Compound Downsampling via Command-Line Interface Demo

Integration with the Joinem DataFrame CLI engine provides direct access to most library functionality (Moreno, 2024). GitHub Container Registry releases allow zero-install execution via Singularity (Kurtzer et al., 2017). The following example shows a compound downsample operation combining canopy and lineage masks.

```
ls -1 "input.csv" `# input path, in alife standard format` \
| singularity exec docker://ghcr.io/mmore500/phyloframe:v0.9.0 `# container` \
  python3 -m phyloframe.legacy._alifestd_pipe_unary_ops `# apply ops in turn` \
  --op "lambda df: pfl.alifestd_mark_sample_tips_canopy_asexual(" \
                        "df, n_sample=5, mark_as='keep_canopy')" \
  --op "lambda df: pfl.alifestd_mark_sample_tips_lineage_asexual(" \
                        "df, n_sample=5, mark_as='keep_lineage')" \
  --op "lambda df: df.assign(extant=df['keep_canopy'] | df['keep_lineage'])" \
  --op "pfl.alifestd_prune_extinct_lineages_asexual" \
  "output.parquet"  # output path
```

A full CLI listing is available via `python3 -m phyloframe --help`.